\documentclass[%
 aip,
 sd,%
 amsmath,amssymb,
reprint,%
]{revtex4-1}
\usepackage[utf8]{inputenc} 
\usepackage[T1]{fontenc} 
\usepackage{amsmath}
\usepackage{url}
\usepackage{physics}
\usepackage{graphicx}
\usepackage{dcolumn}
\usepackage{bm}
\usepackage{ulem}
 \usepackage{cancel}
 \usepackage{array,tabularx,xcolor}
\begin{document}

\preprint{AIP/123-QED}

\title{Accurate measurement of a 96\% input coupling into a cavity using polarization tomography}% Force line breaks with \\
%\thanks{Footnote to title of article.}

\author{P. Hilaire}
 \email{paul.hilaire@c2n.upsaclay.fr}
\affiliation{C2N, Center of Nanosciences and Nanotechnology, Universit\'e Paris-Saclay, Route de Nozay, 91460, France}
 \affiliation{Universit\'e Paris Diderot - Paris 7, 75205 Paris CEDEX 13, France}

\author{C. Ant\'on}
\affiliation{C2N, Center of Nanosciences and Nanotechnology, Universit\'e Paris-Saclay, Route de Nozay, 91460, France}

\author{C. Kessler}
\affiliation{C2N, Center of Nanosciences and Nanotechnology, Universit\'e Paris-Saclay, Route de Nozay, 91460, France}

\author{A. Lema\^itre}
\affiliation{C2N, Center of Nanosciences and Nanotechnology, Universit\'e Paris-Saclay, Route de Nozay, 91460, France}

\author{I. Sagnes}
\affiliation{C2N, Center of Nanosciences and Nanotechnology, Universit\'e Paris-Saclay, Route de Nozay, 91460, France}

\author{P. Senellart}
\affiliation{C2N, Center of Nanosciences and Nanotechnology, Universit\'e Paris-Saclay, Route de Nozay, 91460, France}
%\affiliation{ D\'epartement de physique, Ecole Polytechnique, Universit\'e Paris-Saclay, F-91128 Palaiseau, France}%

\author{L. Lanco}%
\affiliation{C2N, Center of Nanosciences and Nanotechnology, Universit\'e Paris-Saclay, Route de Nozay, 91460, France}
 \affiliation{Universit\'e Paris Diderot - Paris 7, 75205 Paris CEDEX 13, France}

\date{\today}

\begin{abstract}
Pillar microcavities are excellent light-matter interfaces providing an electromagnetic confinement in small mode volumes with high quality factors. They also allow the efficient injection and extraction of photons, into and from the cavity, with potentially near-unity input and output-coupling efficiencies. Optimizing the input and output coupling is essential, in particular, in the development of solid-state quantum networks where artificial atoms are manipulated with single incoming photons.
Here we propose a technique to accurately measure input and output coupling efficiencies using polarization tomography of the light reflected by the cavity.
We use the residual birefringence of pillar microcavities to distinguish the light coupled to the cavity from the uncoupled light: the former participates to rotating the polarization of the reflected beam, while the latter decreases the polarization purity.
Applying this technique to a micropillar cavity, we measure a $53 \pm2 \% $ output coupling and a $96 \pm 1\%$ input coupling with unprecedented precision. 

\end{abstract}

\maketitle

 To enhance the light-matter coupling, optical microcavities \cite{vahala2003optical} have been used to confine the electromagnetic field locally in small mode volumes. Among other solid-state cavities (photonic crystals \cite{Painter1999},
 microdisks \cite{Michler2000,Gayral1999})
 micropillars \cite{Pelton2002} are excellent candidates for light matter interfacing: they provide high quality factors for small mode volumes, potentially low optical losses and high input and output coupling efficiencies. \cite{Loo2010,Arnold2014,Schlehahn2016,Strauf2007}
These structures have already commercial applications thanks to VCSELs \cite{Wiersig2006} which are used, for instance, for optical fibre data transmission and laser reading/writing beams in DVD players. They are also used as efficient exciton-photon interfaces (such as polaritons \cite{gutbrod1998weak,bajoni2008polariton})
and photon-phonon interfaces for high frequency phonons \cite{Fainstein2013,Anguiano2017}.

A major potential application of micropillar devices lies in the development of  a future
quantum photonic network, which requires highly efficient interfaces between photons and artificial atoms.
In this framework, a single photon should ideally couple and deterministically interact with a  single artificial atom such as a semiconductor quantum dot. 
In the context of cavity QED with quantum dots, these devices have already allowed enhancing the spontaneous emission into the cavity mode\cite{G?rard1998} and achieving the exciton-photon strong coupling regime \cite{Reithmaier2004}.
They allow the efficient extraction of indistinguishable single photons \cite{Ates2009,Somaschi2016,Ding2016} for quantum optics applications
and they can potentially be used as a spin-photon interface for quantum computing \cite{Young2011}.
They have recently been used to filter a single photon from an optical pulse, thanks to an optical non linearity at the single photon level \cite{Loo2012b,DeSantis2017}.
This required in particular an excellent input coupling for the incoming photons \cite{Rakher2009b},
 significantly better than the state of the art in photonic crystals.

To perform as efficient interfaces, micropillar cavities must not only confine the electromagnetic field, but also allow the efficient injection/collection of every photon into and from the micropillar cavity, via a careful optical alignment.
In order to obtain an optimal input coupling, a free space gaussian beam must be matched both spatially and spectrally with the confined cavity field.  However, the experimental measurement of this mode matching is evaluated  by analyzing the far-field spatial profile of the input and cavity mode with an uncertainty higher than $5\%$ \cite{Loo2012b}. 

In this Letter, we present a technique to measure accurately the input and  output coupling efficiencies of a micropillar cavity. 
We reconstruct the polarization density matrix of the light reflected by the micropillar cavity using polarization tomography measurements \cite{Anton2017}.
 This allows distinguishing between a  pure polarization state and a general mixed polarization, and, as a consequence, identifying the respective contribution of the light coupled and uncoupled to the cavity.
A record precision is obtained in the measurement of a high input coupling at $\eta_{in}=96\pm1\%$ and of a moderate 
output coupling (defined as the probability for a cavity photon to escape through the top mirror): $\eta_{out}=53 \pm 2\%$.

\begin{figure}
	\centering
		\includegraphics[width=0.50\textwidth]{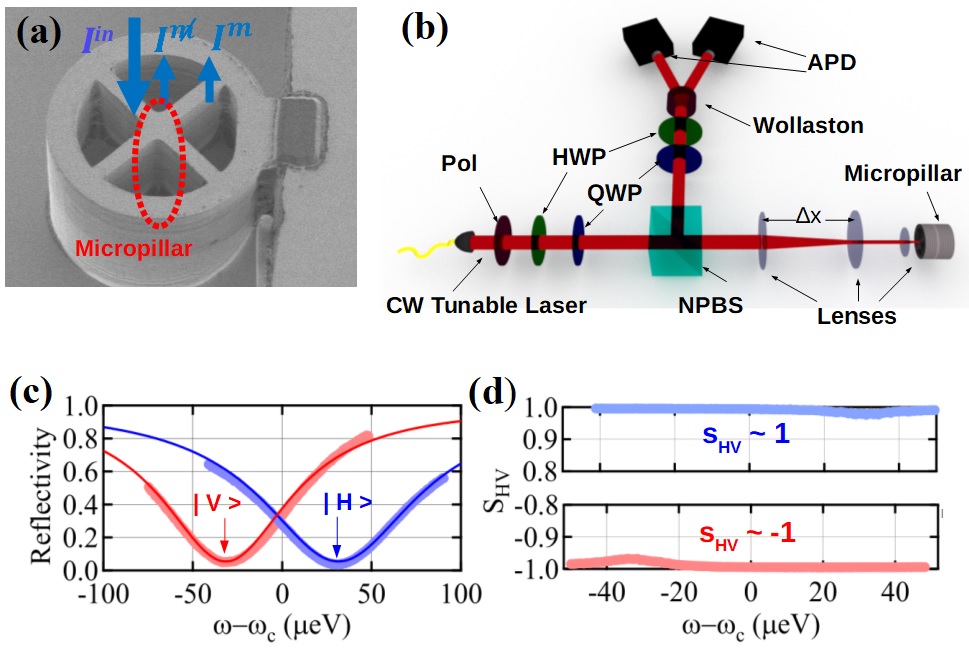}
		\caption{(a)
		SEM image of an electrically-contacted microcavity, where  the central region confines an optical mode and is denoted the micropillar.
		A coherent incident laser with intensity $I^{in}$ is either coupled to the cavity mode and then reflected with an intensity $I^{m}$ or not coupled and then totally reflected with intensity $I^{\not{m}}$.
		(b) Polarization tomography setup: a CW-tunable laser is coupled to a micropillar cavity thanks to a beam adapter ( with adjustable lens separation $\Delta$x). The reflected light  is analysed in polarization with  waveplates and a Wollaston prism. Pol (Polarizer), QWP (Quarter WavePlate), HWP (Half WavePlate), APD (Avalanche Photodiodes), NPBS (Non Polarizing BeamSplitter). (c) Total reflectivity as a function of the laser-cavity detuning $\omega -\omega_c$ for the eigen polarizations of the cavity (vertical in red, horizontal in blue). Points are experimental data while solid lines are theoretical predictions. (d)  Horizontal/Vertical Stokes parameter $s_{HV}$ as a function of the laser-cavity detuning $\omega -\omega_c$ for vertical (in red) and horizontal (in blue) incoming polarizations. For horizontal (resp. vertical) incident polarization, the Stokes parameter $s_{HV}$ remains close to $ +1$ (resp. $-1$), indicating a negligible rotation of polarization. Points are experimental data.}
		\label{fig:ArticleAPLFigureSetup}
\end{figure}

In this experiment, we use a pillar microcavity \cite{nowak2014deterministic,Somaschi2016} (see Fig. \ref{fig:ArticleAPLFigureSetup}(a) ) consisting in a $\lambda$-GaAs cavity, positioned between two distributed Bragg reflectors, with 20 (30) pairs of alternating quarter-wavelength thick $GaAs/Al_{0.9}Ga_{0.1}As$ layers for the top (bottom) mirror. The difference in refractive index between  GaAs and vacuum leads to a lateral confinement between the four ridges of the cavity in the central region, hereafter denoted micropillar. 

The experimental setup is presented in Fig. \ref{fig:ArticleAPLFigureSetup}(b): the pillar microcavity is kept inside a helium gas exchange cryostat at approximately 10 K. The cavity is excited by a tunable continuous wave laser with 1 MHz spectral linewidth.  The polarisation of the incident light is set with a polarizer and waveplates controlling an adjustable polarization state $\ket{\Psi_{in}}$.
The spatial shape of the free space incoming beam is controlled with a beam adapter in order to focus it into the micropillar surface. Therefore, we can adapt the coupling efficiency  $\eta_{in}$ of the external field to the cavity mode.
Given a total incident light intensity  $I^{in}$, the intensity that couples to the micropillar cavity mode is $\eta_{in} I^{in}$; the intensity corresponding to uncoupled light is $(1-\eta_{in}) I^{in}$. In the following, we denote $I^{m}$ the reflected intensity associated to light coupled into the mode, and $I^{\not{m}}$ the reflected intensity associated to light that was not coupled into this mode, as illustrated in Fig. \ref{fig:ArticleAPLFigureSetup}(a). The coupled light is reflected with reflectivity coefficient denoted $R_m$, so that $I^{m}=\eta_{in} R_m I^{in}$.
 The uncoupled light is entirely reflected as long as the focused incident beam is smaller than the micropillar surface, so that $I^{\not{m}}=(1-\eta_{in}) I^{in}$. It will be the case in the following experiments.
    
To perform a complete polarization tomography as also illustrated in Fig. \ref{fig:ArticleAPLFigureSetup}(b), the reflected beam is separated in two orthogonally-polarized components in various polarization bases, using calibrated waveplates and a Wollaston polarizing prism. The input and output field intensities are measured with avalanche photodiodes. By adjusting the waveplates of the polarization analyzer, we measure the intensities $I_H$ and $I_V$ in the horizontal/vertical polarization basis, the intensities $I_D$ and $I_A$ in the diagonal/anti-diagonal polarization basis, and $I_R$ and $I_L$ in the right-handed/left-handed circular polarization basis. 
For a given set of orthogonally polarized intensities  $I_{\|} / I_{\bot}$ , 
 we define the corresponding Stokes component as  $s_{\| \bot}=(I_{\|}-I_{\bot})/(I_{\|} + I_{\bot})$. 
This method \cite{Anton2017} allows measuring the density matrix of the polarization state, and representing it in the
 Poincar\'e sphere as a vector with coordinates of
 $s_{HV}$, $s_{DA}$ and $s_{RL}$ , ranging between $-1$ and $+1$. 
The norm of the Poincar\'e vector $\sqrt{s_{HV}^2+s_{DA}^2+s_{RL}^2}$ is the purity of the polarization density matrix, equal to 1 for a pure polarization state.
Thus we are able to reconstruct the polarization density matrix of the reflected light and represent it in the Poincar\'e sphere.

The fundamental mode energy of the cavity is $\omega_c=1.3365$ $\mu eV$; it is splitted due to a small geometrical ellipticity, leading to linearly polarized horizontal (H) and vertical (V) modes.
We first excite the device with horizontally (respectively vertically) polarized light $\ket{\Psi_{in}}=\ket{H}$ ($\ket{V}$). The blue (red) curve in Fig. \ref{fig:ArticleAPLFigureSetup}(c) displays the total reflectivity as a function of the laser-cavity detuning, evidencing a cavity mode splitting $\delta\omega=63 \pm 1$ $\mu eV$.With this notation, the resonance energies are given by $\omega_H = \omega_c + \delta\omega/2$ and $\omega_V = \omega_c - \delta\omega/2$ for the horizontally-polarized and vertically-polarized cavity modes, respectively. 
In this figure as in the following ones, the experimental points are compared to a theoretical fit that will be described later on. The reflectivity curve presents a Lorentzian reflectivity dip with linewidth $\kappa_H=105 \pm 5$ $\mu eV$ ($\kappa_V=86 \pm 5$ $\mu eV$). We can also see that the Lorentzian dip does not reach zero reflectivity value, as would be obtained for a perfect input coupling and a top mirror output coupling of $50\%$. Indeed, the reflectivity coefficient for each mode $i=\{H,V\}$ is \cite{Walls2008}:

\begin{equation}
 r_i=1-2\eta_{out}/(1-2i(\omega-\omega_{i})/\kappa_i)
\end{equation}

So the reflectivity of the mode is $R_{m,i}=|r_i|^2$ where $\omega_{i}$ is the cavity mode energy and $\kappa_i$ the cavity linewidth: it can go down to zero at $\omega=\omega_i$ if $\eta_{out}=50 \%$.
In general, however, the total reflectivity $R_{tot,i}$ takes into account both coupled and uncoupled reflected light:
 $R_{tot,i}=(I^m+I^{\not{m}})/ I^{in}=(1-\eta_{in})+\eta_{in} R_{m,i}$. 
As the reflectivity values provides a single constraint for two unknown variables, we cannot deduce the contributions of $\eta_{in}$ and $\eta_{out}$ unambiguously without complementary information. 
For example, in the case of Fig. \ref{fig:ArticleAPLFigureSetup}(c), the minimal reflectivity of $5\%$ can be explained by $\eta_{in}=95 \%$ and $\eta_{out}=50 \%$ or by $\eta_{in}=100 \%$ and $\eta_{out}=60 \%$ (or $40 \%$). We can only deduce that $\eta_{in} \in [95 \% ; 100 \%]$ and $\eta_{out} \in [40\% ; 60 \%]$.

For the same incoming polarization, $\ket{\Psi_{in}}=\ket{H}$ (resp. $\ket{V}$) and for each laser wavelength, we measure the Stokes parameter $s_{HV}$ for the reflected light as a function of $\omega-\omega_c$ (Fig. \ref{fig:ArticleAPLFigureSetup}(d)). We obtain for all wavelengths that $s_{HV} \approx 1$ (resp. $s_{HV} \approx -1$) showing that $I_H \gg I_V$ (resp. $I_H \ll I_V$). This shows that given an incident polarization $\ket{\Psi_{in}}=\ket{H}$ or $\ket{V}$, the reflected light is also a pure polarization state $\ket{\Psi_{m}} \approx\ket{\Psi_{in}}$.

\begin{figure}
	\centering
		\includegraphics[width=0.45\textwidth]{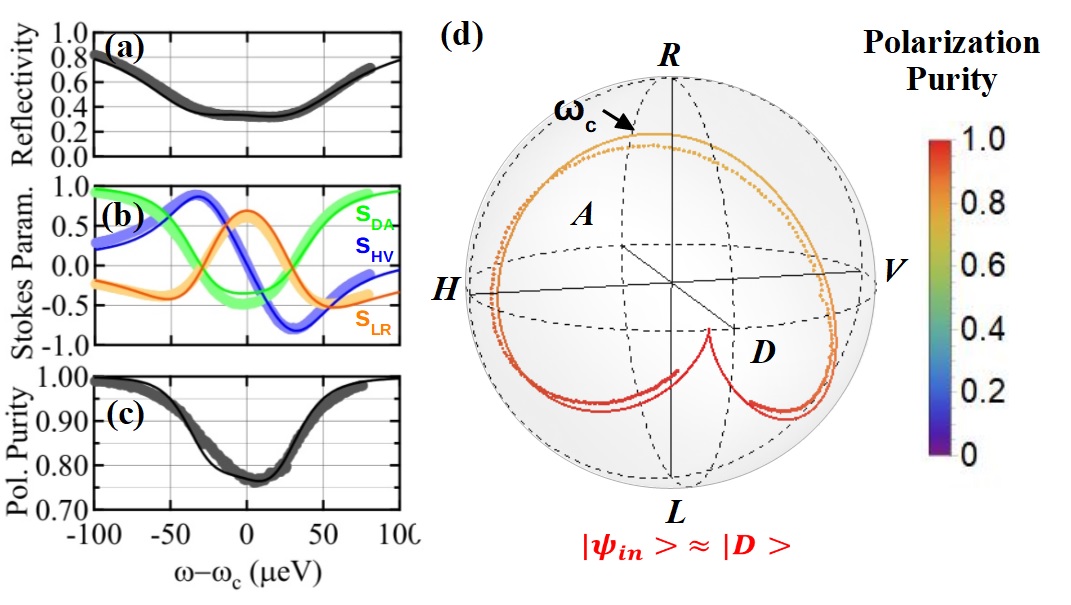}
		\caption{
		(a) Total reflectivity ,   (b) Stokes parameters  ($S_{HV}$ in blue/ $S_{DA}$ in green / $S_{LR}$ in Orange) and (c) polarization purity  as a function of the laser-cavity detuning $\omega -\omega_c$ for diagonally polarized incident polarization ($\ket{\Psi_{in}} \approx \ket{D}$).  (d) From the Stokes parameters, we plot the polarization density matrix for each laser photon energy. The points are the experimental data while the line is the theoretical prediction. The polarization purity is indicated in color scale in the Poincar\'e sphere.  $\omega_c$ denotes the point corresponding to the laser-cavity resonance. 
		}		                                                                    
	\label{fig:ArticleAPLFigureReflStokes}
	
\end{figure}

A strikingly different situation appears if we excite the cavity with a diagonal polarization $\ket{\Psi_{in}}\approx \ket{D}=(\ket{H}+\ket{V})/\sqrt{2}$, away from its cavity polarization axis: we obtain the reflectivity curve displayed in Fig. \ref{fig:ArticleAPLFigureReflStokes}(a) and the complete set of Stokes parameters $s_{HV}$,$s_{DA}$ and $s_{LR}$ described in Fig. \ref{fig:ArticleAPLFigureReflStokes}(b). In the absence of uncoupled light, i. e. $\eta_{in}=1$, the total reflectivity would be described by $R_m=(|r_H|^2+|r_V|^2)/2$ and the reflected polarization state by $\ket{\Psi_{m}}= (r_H \ket{H}+r_V \ket{V})/(|r_H|^2+|r_V|^2)$, rotated from $\ket{\Psi_{in}}$ as $r_H(\omega) \neq r_V(\omega)$. A polarization rotation is indeed observed, as shown by the variation of the Stokes parameters $s_{HV}$, $s_{DA}$ and $s_{LR}$ as function of the input laser energy. However, the reflected output is not a pure polarization state, as shown in Fig. \ref{fig:ArticleAPLFigureReflStokes}(c), displaying a polarization purity below unity when $\omega \approx \omega_c$. This behavior is also complementarily illustrated in Fig. \ref{fig:ArticleAPLFigureReflStokes}(d), where the polarization density matrix is displayed in the Poincar\'e sphere for various values of $\omega$. When $\omega$ is detuned from $\omega_c$, the reflected polarization is close to the incoming one ($\ket{\Psi_{m}} \approx \ket{\Psi_{in}}$), but notably different  when $\omega \approx \omega_c$. In Fig. \ref{fig:ArticleAPLFigureReflStokes}(d), the experimental points are encoded in a colorscale representing the corresponding polarization purity, which decreases down to $ 76 \%$ when $\omega \approx \omega_c$. Therefore, the reflected light polarization can not be considered as a pure polarization state.

To explain this depolarization, we have to take into account the limited input coupling $\eta_{in}$, which is the overlap between the spatial profiles of the incoming beam and the fundamental cavity mode.
The light coupled to the  cavity is indeed reflected in a pure polarization state $\ket{\Psi_{m}} \sim \alpha\ket{H} + \beta \ket{V}$ with a reflected intensity $I^m =\eta_{in} R_m I^{in}$. The uncoupled light is fully reflected with an unrotated polarization state $\ket{\Psi_{in}}$,
whose reflected intensity is $I^{\not{m}}=(1-\eta_{in})I^{in}$ (see Fig. \ref{fig:ArticleAPLFigureSetup} (a)). As an example, for $\ket{\Psi_{in}}=\ket{D}$, $R_m$ is given by $(|r_H|^2+|r_V|^2)/2$ and $\ket{\Psi_{m}}=(r_H \ket{H}+r_V \ket{V} )/\sqrt{|r_H|^2+|r_V|^2}$.

An important property of the coupled and uncoupled components $I^m$ and $I^{\not{m}}$ is that the corresponding optical beams have orthogonal spatial profiles, and thus their superposition does not lead to interference in the total intensity. Regarding the coupled component, its spatial profile is governed by the spatial shape of the fundamental cavity mode only. The spatial profile of the uncoupled component arises from the contribution of other modes, which all have spatial profiles orthogonal to that of the fundamental one. The intensities of the two beams thus sum up without interference: this leads to a total reflectivity $(I^m+I^{\not{m}} )/I^{in}$ and to a polarization density matrix given by:

\begin{equation}
\rho=p \ket{\Psi_{m}} \bra{\Psi_{m}} +(1-p) \ket{\Psi_{in}} \bra{\Psi_{in}}
\label{eq}  
\end{equation}
with 

\begin{equation}
p=\frac{I^m}{I^m + I^{\not{m}}}=\frac{\eta_{in} R_m}{(1-\eta_{in})+\eta_{in} R_m}\;
\end{equation}

\begin{figure}
	\centering
		\includegraphics[width=0.50\textwidth]{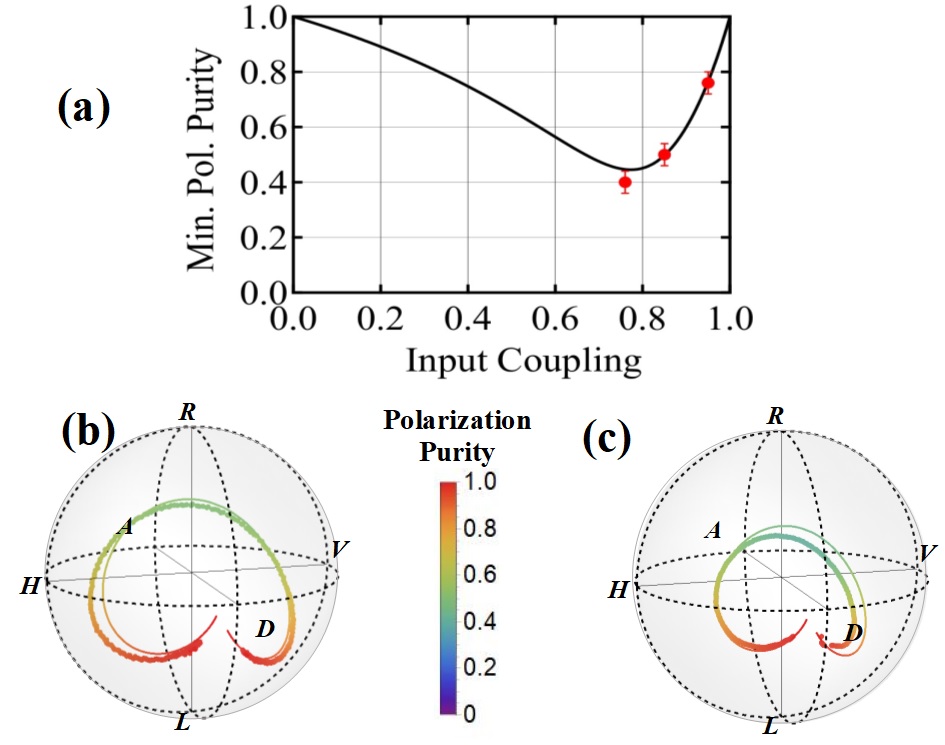}
		\caption{(a) The theoretical minimum of polarization is plotted as a function of the input coupling with the 3 experimental minima measured.
		The Poincar\'e sphere (b) ( respectively (c)) was obtained by scanning in wavelength the cavity resonance for an estimated coupling $ \eta_{in}=0.85$ (respectively $0.76$).The points are the experimental data while the line is the theoretical prediction. The polarization purity is indicated in color scale in the Poincar\'e sphere. }
	\label{fig:ArticleAPLFigurePoincareSphere}
	
\end{figure}

The theoretical fits displayed by solid lines in Figs. \ref{fig:ArticleAPLFigureSetup}-\ref{fig:ArticleAPLFigurePoincareSphere} are obtained with this model,  where $\ket{\Psi_{in}}= \ket{H}$ or $\ket{V}$ in Fig. \ref{fig:ArticleAPLFigureSetup} and $\ket{\Psi_{in}} \approx \ket{D}$ in Figs. \ref{fig:ArticleAPLFigureReflStokes} and \ref{fig:ArticleAPLFigurePoincareSphere}.

In the latter case, the reflected light has two contributions ($I^{m}$ and $I^{\not{m}}$) which incoherently superpose different polarizations, as $ \ket{\Psi_{m}} \neq \ket{\Psi_{in}}$ when $\omega \approx \omega_c$. 
 In this case, the polarization purity is below unity as shown in Fig. \ref{fig:ArticleAPLFigureReflStokes}(c). However, far from the cavity resonance,  $r_H \approx r_V \approx 1$ so that $\ket{\Psi_{m}} \approx \ket{\Psi_{in}}$: thus $\rho=\ket{\Psi_{in}}\bra{\Psi_{in}}$, corresponding to an unrotated and pure polarization state.

As we now describe, thanks to the polarization tomography technique and to the theoretical model presented above, the input and output couplings can be measured accurately.
Fig. \ref{fig:ArticleAPLFigurePoincareSphere}(a) displays the dependence of the minimum polarization purity achieved at $\omega \approx \omega_c$ as a function of the input coupling.
For high values of input coupling (i. e. $\eta_{in}>0.8$), the minimum  polarization purity  is very sensitive to small changes in $\eta_{in}$. 
From Fig. \ref{fig:ArticleAPLFigurePoincareSphere}(a), we  see that for a given value of the minimal polarization purity, there are actually two possible values of  $\eta_{in}$. However, it is easy to discriminate which is the correct value of $\eta_{in}$ by looking at the  amplitude of the polarization rotation induced in the Poincar\'e sphere (see Fig. \ref{fig:ArticleAPLFigureReflStokes}(c)): for a low $\eta_{in}$, most of the light is uncoupled and experiences no rotation, for a high  $\eta_{in}$, most of the light experiences a rotation of polarization. By fitting  the reflectivity curves in Fig. \ref{fig:ArticleAPLFigureSetup}(b),  Fig. \ref{fig:ArticleAPLFigureReflStokes}(a), and the polarization density matrix (which corresponds to the Stokes parameters in Fig. \ref{fig:ArticleAPLFigureReflStokes}(b) and the Polarization purity in Fig. \ref{fig:ArticleAPLFigureReflStokes}(c) or the Poincar\'e sphere in Fig. \ref{fig:ArticleAPLFigureReflStokes} (d), we can accurately and unambiguously estimate the value of the top mirror output coupling $\eta_{out}=53 \pm 2\%$ and the input coupling $\eta_{in}=96 \pm 1\%$. Such a value corresponds to the best spatial overlap that we could experimentally achieve, by careful optical alignment, between the incoming free space optical beam and the spatial profile of the fundamental cavity mode. This overlap could potentially be improved up to 100\%, by further shaping the incoming optical beam to exactly match the cavity mode profile.

Furthermore, by modifying the spatial size of the free-space incoming beam with the beam adapter ($\Delta x$, see Fig. \ref{fig:ArticleAPLFigureSetup} (a)),
 we can vary $\eta_{in}$. As illustrated in Figs. \ref{fig:ArticleAPLFigurePoincareSphere}(b,c)
 this was performed for two other incident beam sizes and thus different experimental values of $\eta_{in}$. Figures \ref{fig:ArticleAPLFigurePoincareSphere}(b,c) display the reconstructed polarization density matrix of the 
reflected light, as in Fig. \ref{fig:ArticleAPLFigureReflStokes}(d), with the same experimental conditions and the same
 input polarization $\ket{\Psi_{in}} \approx \ket{D}$. 
In order to fit the data, we use the same cavity linewidths and output coupling $\eta_{out}=53 \pm2 \%$, which was determined thanks to the previous experiment and to the theoretical model of Eq. \ref{eq}. The only parameter that is varied to fit the Poincar\'e sphere displayed in Fig. \ref{fig:ArticleAPLFigurePoincareSphere}(b) and (c) is $\eta_{in}$. The agreement between theory and experiment allows us to estimate a mode matching of $85 \pm 2\%$ and $76 \pm 3\%$ respectively.

In summary, we have demonstrated that polarization tomography is a robust technique to determine input and output couplings with high accuracy. Indeed, the polarization purity of the reflected light is very sensitive to a slight amount of uncoupled light. In addition, we use a simple theoretical model to interpret the data and accurately determine an experimental input coupling of 
$96\pm 1 \%$ and an output coupling of $53 \pm2\%$. 
In the context of photonic quantum networks, 
such high input couplings are crucial to the realization of deterministic photon-photon gates \cite{Koshino2010,Shomroni2014}, where a first photon  must be coupled to the cavity to modify the state of a second one. %with high accuracy

This work is supported by the Agence Nationale de la Recherche (ANR) (ANR-
458 12-BS10-0010, ANR-14-CE32-0012, and the "Investissements d'avenir" program Labex NanoSaclay ANR-10-LABX-0035); H2020 European
459 Research Council (ERC) Starting Grant QD-CQED (277885);
460 H2020 Marie Skłodowska-Curie Actions (MSCA) Marie Curie
461 Fellowship SQuaPh (702084); Labex (NanoSaclay).

\end{document}